\documentclass{llncs}
\usepackage{etex}
\usepackage{url}
\usepackage[figuresright]{rotating}
\usepackage[lofdepth,lotdepth]{subfig}

\usepackage{oz}

\usepackage{algpseudocode}
\algrenewcommand{\algorithmiccomment}[1]{\hfill\(/\!\!/\) #1}
\algrenewcommand\algorithmicfunction{\textbf{operation}}
\algnewcommand\algorithmicshared{\textbf{Shared:}}
\algnewcommand\Shared{\item[\algorithmicshared]}
\algnewcommand\algorithmictype{\textbf{Type:}}
\algnewcommand\Type{\item[\algorithmictype]}
\algnewcommand\algorithmicatomic{\textbf{atomic}}
\algdef{SE}[ATOMIC]{Atomic}{EndAtomic}{\algorithmicatomic}{\algorithmicend\ \algorithmicatomic}
\algloopdefx{Either}{\textbf{either}}
\algcloopdefx{Either}{Or}{\textbf{or}}
\algloopdefx{When}[1]{\textbf{when} #1}
\renewcommand{\gets}{\mathbin{:=}}

\newcommand{\psvar}[1]{\ensuremath{\mathit{#1}}}

\newcommand{\psterm}[1]{\ensuremath{\mathrm{#1}}}

\newcommand{\ph}{\ensuremath{\varphi}}
\newcommand{\tuple}[1]{\ensuremath{\left< #1 \right>}}
\newcommand{\half}{\ensuremath{\frac{1}{2}}}

\newcommand{\Preds}{\ensuremath{\mathcal{P}}}
\newcommand{\A}{\ensuremath{\mathcal{A}}}
\newcommand{\pp}[1]{\ensuremath{\mathsf{#1}}}
\newcommand{\ppp}{\pp{p}}

\newcommand{\peq}{\pp{eq}}

\newcommand{\pnext}{\pp{next}}

\newcommand{\px}{\pp{x}}
\newcommand{\pa}{\pp{a}}
\newcommand{\pb}{\pp{b}}

\newcommand{\pis}[1]{\pp{is\_{#1}}}

\newcommand{\phas}[1]{\pp{has[#1]}}

\newcommand{\psucc}[1]{\pp{succ[#1]}}

\newcommand{\pat}[1]{\pp{at[#1]}}

\newcommand{\pif}[1]{\pp{if[#1]}}
\newcommand{\pnif}[1]{\pp{nif[#1]}}

\renewcommand{\imp}{\rightarrow}     

\usepackage{tikz}
\usetikzlibrary{arrows,shapes,positioning,fit,decorations.pathmorphing} 
\tikzset{auto,node distance=1.5cm}
\tikzset{every node/.style={font=\small}}

\tikzset{state border/.style={draw,rectangle,rounded corners,inner sep=.5cm}}
\tikzset{state name/.style={font=\footnotesize}}
\tikzset{state transform/.style={->,thick,double,decorate,decoration={snake,post length=.2cm}}}
\tikzset{state transform label/.style={font=\Large}}

\tikzset{object/.style={draw,thick,align=center,font=\small}}
\tikzset{summary/.style={double}}
\tikzset{nodeCA/.style={object,shape=rectangle,minimum size=1.6cm}}
\tikzset{datum/.style={object,shape=circle,minimum size=1cm}}
\tikzset{thread/.style={object,shape=regular polygon,regular polygon sides=6,inner sep=-.1cm}}
\tikzset{nodesCA/.style={nodeCA,summary}}
\tikzset{data/.style={datum,summary}}
\tikzset{threads/.style={thread,summary}}
\tikzset{tight/.style={inner sep=0}} 

\tikzset{bpred/.style={->,thick,shorten <=2pt,shorten >=2pt,}}
\tikzset{true/.style={bpred}}
\tikzset{maybe/.style={bpred,dotted}}
\tikzset{loop left/.style={loop,left,out=215,in=145,distance=1.2cm}}
\tikzset{loop right/.style={loop,right,out=-35,in=35,distance=1.2cm}}
\tikzset{loop above/.style={loop,above,out=55,in=125,distance=1.2cm}}
\tikzset{loop below/.style={loop,below,out=305,in=235,distance=1.2cm}}

\usepackage{color}

\begin{document}
\title{Collapsing Threads Safely with Soft Invariants}
\author{David Friggens\inst{1} \and Lindsay Groves\inst{2}}
\institute{
\email{david@frigge.nz}
\and
Victoria University of Wellington\\
\email{lindsay@ecs.vuw.ac.nz}}

\maketitle

\begin{abstract}
Canonical abstraction is a static analysis technique that represents states as 3-valued
logical structures, and produces finite abstract systems.
Despite providing a finite bound, these abstractions may still suffer from the state
explosion problem. 
Notably, for concurrent programs with arbitrary interleaving, if threads in a state are
abstracted based on their location, then the number of locations will be a combinatorial
factor in the size of the statespace.

We present an approach using canonical abstraction that avoids this state explosion by
``collapsing'' all of the threads in a state into a single abstract representative. Properties
of threads that would be lost by the abstraction, but are needed for verification, are
retained by defining conditional ``soft invariant'' instrumentation predicates.
This technique is used to adapt previous models for verifying linearizability of nonblocking concurrent
data structure algorithms, resulting in exponentially smaller statespaces.
\end{abstract}

\section{Introduction}
\label{sec:intro}

Canonical abstraction \cite{sagi:para02} is a powerful technique for reasoning about heap-manipulating programs.
It is effective for analyzing concurrent systems with interleaving threads, 
allowing on-the-fly model checking of safety properties.
However, the standard abstraction is still affected by the statespace explosion problem.
In this paper, we introduce a way of using canonical abstraction for concurrent systems that is
combinatorially more efficient.

\subsection{Motivation}
\label{sec:intro:motivation}

Canonical abstraction represents states as 3-valued logical structures.
A (potentially infinite) concrete statespace is abstracted to a finite abstract statespace by mapping the
(potentially infinite) heap objects in each state to a finite number of abstract heap objects.
The abstract heap objects are equivalence classes with respect to a predetermined (finite) set of unary
predicates. Each abstract heap object represents all of the heap objects in the concrete state with the
same truth valuation of the abstraction predicates.

Canonical abstraction was initially developed for sequential systems, but
Yahav and Sagiv \cite{yaha:veri10} showed how it can be used for concurrent systems
with interleaving threads.
The logical structures are extended to include an object for each thread, and predicates
for the possible program locations (labels) of the threads.
The set of abstraction predicates includes the location predicates
and possibly properties of thread fields, such as whether a particular field is null.
As a result, the number of equivalence classes of abstract thread objects is determined
by the number of locations and possibly some user defined properties.

This approach has some benefits. For example, if an important property of a system is that the
\(Y\) field of a thread is always null at location \(X\), then this is implicitly recorded in the abstract
statespace. In every reachable abstract state there will only be a thread object with the
``at location \(X\)'' predicate true and with the ``\(Y\) is null'' predicate true, never with ``\(Y\) is null'' false.

There is an important downside to this approach, however.
For each abstract heap configuration, there is an abstract state with a different combination of
interleaving thread states. For example, if there are three interleaving locations \(X\), \(Y\) and \(Z\),
then every abstract heap configuration has seven corresponding abstract states: three when all
threads are at the same location (\(X\), \(Y\) or \(Z\)), three when all threads are at one of two
locations (\(\{X,Y\}\), \(\{X,Z\}\) or \(\{Y,Z\}\)), and one when there are threads at all three locations.

As the number of possible interleaving thread states increases, the number of these combinations
increases, and the less efficient the abstract statespace becomes.
Can we avoid the statespace explosion, but still make the abstraction precise enough to verify
interesting properties?

\subsection{Soft Invariants}
\label{sec:intro:si}

The ``obvious'' fix is to map all thread objects to a single abstract thread object in each abstract state.
This avoids the problem of combinatorial thread objects, but loses too much precision --- for example,
the abstract thread object can record whether \(Y\) is null for all threads, whatever location they are in,
but not for only threads at location \(X\).

We propose new derived unary predicates to explicitly preserve certain properties of threads at
particular locations. These predicates are defined using conditional formulas of the form, e.g.\ ``if
the location is \(X\) then the field \(Y\) is null''.
The predicate will be true for all threads at locations other than \(X\), and if \(Y\) is always null
at location \(X\), then the predicate will be true for all those threads also.
Thus the predicate will be true for the single abstract thread object in each abstract state.

If \(Y\) is sometimes non-null at location \(X\), then the predicate will be false for those thread
objects, and will be ``unknown'' for the resulting abstract thread object.\footnote{%
In the 3-valued semantics, ``unknown'' indicates that the abstract predicate represents
instances in corresponding concrete states when it is true, and when it is false.}
We call these predicates ``soft invariants'', as they allow us to explicitly preserve invariants
of the system (such as ``at location \(X\), field \(Y\) is always null'') but without having to prove that 
the property is invariant. If they are not actually invariant then the predicate does not affect the analysis.


The paper is structured as follows.
In Section~\ref{sec:bg}, we give some background details about canonical abstraction.
In Section~\ref{sec:ex}, we illustrate the use and effects of soft invariants with a simple
example, advancing a pointer along a circular linked list.
In Section~\ref{sec:results}, we show the results of using soft invariants to verify two
nonblocking data structures, improving on our previous work.
In Section~\ref{sec:choice}, we discuss choosing soft invariants, and demonstrate that
they can be safely over-defined.
Finally, in Section~\ref{sec:relwork} we discuss related work.


\section{Preliminaries}
\label{sec:bg}

\subsubsection{States as logical structures}

Sagiv et~al.~\cite{sagi:para02}
represent states as 3-valued logical structures, where predicates describe relationships
between objects, and multiple concrete
objects can be represented by a single abstract ``summary object''.
Since a summary object can represent two or more concrete objects, an abstract state
with summary objects can represent an infinite number of concrete states.

First, a finite set of predicates \(\Preds=\{\peq, \ppp_1, \dots, \ppp_n\}\) is fixed for the analysis,
and we define \(\Preds_k\) to be the set of \(k\)-ary predicates in \(\Preds\) (the equality predicate
\peq\ has arity 2).
Then, an (\emph{abstract}) \emph{configuration} \(S = \tuple{U,\iota}\)
has a \emph{universe} \(U\) that is a (finite or infinite) set of objects and an
\emph{interpretation} \(\iota\) over the logical
values true ($1$), false ($0$) and unknown ($\half$). For each \(k\)-ary predicate \(\ppp\),
\[ \iota(\ppp) : U^k \to \{ 1,0,\half \} \]
Additionally, for each \(u_1,u_2 \in U\) where \(u_1 \neq u_2\),
\(\iota(\peq)(u_1,u_2)=0\).
An object \(u\), for which \(\iota(\peq)(u,u)=\half\) is called a \emph{summary object}.
A \emph{concrete configuration} is one where all interpretations are 1 or 0.

Intuitively, one configuration represents a less abstract one if it contains the same
information, except for some conservative information loss. In other words, it has the same
universe of objects, though some may have been merged together into summary objects, and
it has the same predicate interpretations, though some may have become unknown.
This is formalised by the notion of embedding, which relates configurations
that are related by conservative information loss.

We say that a configuration \(S_1 = \tuple{U_1,\iota_1}\)
\emph{embeds} into a configuration \(S_2 = \tuple{U_2,\iota_2}\) if there exists
a surjective function \(f : U_1 \to U_2\) such that for every \(k\)-ary predicate \(\ppp\), and
\(u_1,\dots,u_k \in U_1\),
\[ \iota_1(\ppp)(u_1,\dots,u_k) \sqsubseteq \iota_2(\ppp)(f(u_1),\dots,f(u_k)) \]
where, for \(l_1,l_2 \in \{1,0,\half\}\), \(l_1 \sqsubseteq l_2\) iff \(l_1=l_2\) or \(l_2=\half\).

We further define a \emph{tight embedding} to be one that minimises information loss, i.e.\ 
a predicate interpretation only becomes unknown if two objects are being merged together, one
which has a true interpretation and the other a false interpretation.
Formally, there exists a surjective function \(f : U_1 \to U_2\) such that for every \(k\)-ary
predicate \(\ppp\), and \(u_1,\dots,u_k \in U_2\),
\[ \iota_2(\ppp)(u_1,\dots,u_k) = \\ \t1
\left\{\begin{array}{cl}
1 & \mathrm{if~} \all {u_1' \in f^{-1}(u_1)}, \dots, {u_k' \in f^{-1}(u_k)} @ 
\iota_1(\ppp)(u_1',\dots,u_k') = 1 \\
0 & \mathrm{if~} \all {u_1' \in f^{-1}(u_1)}, \dots, {u_k' \in f^{-1}(u_k)} @ 
\iota_1(\ppp)(u_1',\dots,u_k') = 0 \\
\half & \mathrm{otherwise}
\end{array}\right.
\]

\subsubsection{Canonical abstraction}
\emph{Canonical abstraction} is a method for constructing tight embeddings. Given
a subset of the unary predicates \(\A \subs \Preds_1\), called the \emph{abstraction predicates},
we map objects in the original configuration to the same abstract object if they have the same
interpretations over the abstraction predicates. The interpretation in the abstract configuration
is constructed as per the definition of tight embeddings above.
We say that a configuration is canonically abstract, with respect to \(\A\), if it is the
canonical abstraction of itself.

Canonical abstraction has a number of important properties:
\begin{itemize}
\item Every configuration has a single canonical abstraction, as each object has a single
canonical mapping in the embedding function.
\item Since there are a finite number of abstraction predicates, it follows that there is a finite
bound on the number of objects in the universe of a canonically abstract configuration, and
thus a finite bound on the number of potential states in an abstract system.
\end{itemize}

The soundness of the canonical abstraction approach rests upon the Embedding Theorem
\cite[Theorem 4.9]{sagi:para02}.
Informally, this says that if a structure \(S\) embeds into a structure \(S'\), then
any information extracted from \(S'\) via a formula \(\ph\) is a conservative
approximation of the information extracted from \(S\) via \(\ph\).
Alternatively, if we prove a property \(\ph\) true or false in \(S'\), then we know
it has the same value in \(S\).

\subsubsection{Graphical representation} 
It is often helpful for comprehension to represent logical structures using graphs.
We use
graph nodes to represent objects in the universe, with a double line indicating a
summary object.
Labels on nodes represent unary predicates --- for each object, they are present if true,
  present with annotation (e.g.\ ``\(\px=\half\)'') if unknown, or absent if false.
Labelled arrows represent binary predicates --- for each pair of objects, a solid line if true,
  a dotted line if unknown, or absent if false.
Additionally, we use different node shapes to represent object type predicates, instead of labels ---
hexagons for threads and rectangles for linked list nodes.

\subsubsection{Refining abstractions}

Canonical abstraction using the fixed predicates \Preds\ is often too coarse, resulting
in too much information being lost (i.e.\ evaluating to unknown) for a property to be verified. A key method
for refining abstractions is to introduce additional predicates that record properties derived
from the other predicates.
These \emph{instrumentation predicates} add no new information to a concrete state, since they
evaluate to the same truth values as their defining formulas. However, in an abstract state they
may add information: an instrumentation predicate may evaluate to a definite value (true or false)
whilst its defining formula may evaluate to unknown.
Additionally, unary instrumentation predicates are usually included in the set of abstraction predicates \(\A\),
which can prevent some objects from being merged together into summary objects.
In this paper, we make a point of removing certain instrumentation predicates from \(\A\)
so that thread objects \emph{are} merged together.

%

\subsubsection{Tool Support}

Canonical abstraction has been implemented in TVLA\footnote{\url{http://www.cs.tau.ac.il/~tvla/}} 
(for Three Valued Logic Analyzer), produced at Tel Aviv
University \cite{leva:tvla00,bogu:reva07}.
All of the results in this paper use the ``partial join'' option \cite{mane:part04}.

\section{Example}
\label{sec:ex}

Figure~\ref{fig:inc} shows a simple operation to advance a global pointer \psvar{x} along
a linked list. The thread takes a snapshot (\psvar{a}) of \psvar{x}, then of that node's
\psvar{next} field (\psvar{b}). If the snapshot is unchanged, the atomic Compare-and-Swap (CAS)
command\footnote{$\psterm{CAS}(\psvar{x},\psvar{a},\psvar{b})$ is an atomic command that sets
\psvar{x} to \psvar{b} and returns true only if $\psvar{x}=\psvar{a}$; otherwise it just
returns false.} 
advances \psvar{x}. Otherwise, if another thread has altered \psvar{x}, then the loop
is restarted to try again.

\begin{figure}[t]
\begin{center}
\begin{minipage}[t]{28ex}
  \begin{algorithmic}[1]
    \Function{Inc}{\,}
    \Repeat
      \State $\psvar{a} \gets \psvar{x}$
      \State $\psvar{b} \gets \psvar{a}.\psvar{next}$
    \Until{$\psterm{CAS}(\psvar{x},\psvar{a},\psvar{b})$}
    \State \Return 
    \EndFunction
  \end{algorithmic}
\end{minipage}
  \caption{Operation to advance a pointer \psvar{x} along a linked list}
  \label{fig:inc}
\end{center}
\end{figure}

To illustrate the use of soft invariants, we will consider a system composed of an arbitrary
number of threads that repeatedly perform the \textsc{Inc} operation.
The system is initialised with a circular linked list with a single shared variable \psvar{x}
pointing at one of the list nodes.
We will check the safety property that \psvar{x} is never null.

\subsection{Predicates}
\label{sec:ex:pred}

To begin the model, we define unary core predicates
for the object types, global variable, and thread locations:\footnote{%
The square brackets have no meaning other than being a visual indicator.}
\[
  \pis{thread}, \pis{node}, \px,
  \pat{idle}, \pat{line3}, \pat{line4}, \pat{line5}, \pat{line6}
\]
and binary core predicates for the object's fields:
\[
  \pnext, \pa, \pb
\]
To preserve the fact that the nodes' \psvar{next} fields are always non-null, and to record whether
the thread fields are null or not, we define the following unary instrumentation predicates:
\[
  \phas{next}(v) \iff \exi u @ \pnext(v,u)\\
  \phas{a}(v)    \iff \exi u @ \pa(v,u)\\
  \phas{b}(v)    \iff \exi u @ \pb(v,u)
\]
To record whether a thread's \psvar{b} field is the \psvar{next}-successor of its \psvar{a}
field, we define the following unary shape predicate \cite{frig:shap14}:
\[
  \psucc{a,b,next}(v) \iff \exi u_1,u_2 @ \pa(v,u_1) \land \pb(v,u_2) \land \pnext(u_1,u_2)
\]

\subsection{Initial State}
\label{sec:ex:init}

\begin{figure}
\centering
  \begin{tikzpicture}
    \node[threads] (t)  {\pat{idle}};
    \node[nodeCA,right=of t] (nx) {\px\\\phas{next}};
    \node[nodesCA,right=of nx] (ns) {\phas{next}};
    
    \path (nx) edge [maybe,bend left] node {\pnext} (ns)
          (ns) edge [maybe,loop right] node {\pnext} (ns)
          (ns) edge [maybe,bend left] node {\pnext} (nx);
  \end{tikzpicture}
\caption{Initial state of the example system}
\label{fig:init}
\end{figure}
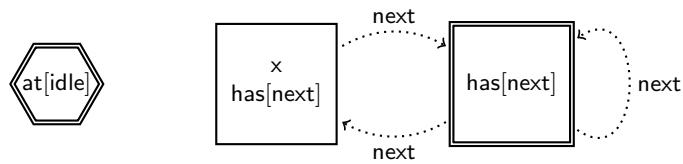

In the initial state, all threads are at the idle location, and all their fields are null.
The node objects are indistinguishable except for the one pointed to by \psvar{x}.
Figure~\ref{fig:init} shows a diagram of the state.

\subsection{Combinatorial States}
\label{sec:ex:yahav}

By default, the set of abstraction predicates is the set of unary predicates.
In this simple example system, the circular list is identical in every abstract state --- when
\psvar{x} advances along the list, the old node is abstracted with the rest
and the abstract representation remains unchanged. Thus the
differences between the abstract states are solely due to the thread objects.

\begin{table*}
   \centering
   \begin{tabular}{cccccccc}
     \pat{idle} & \pat{line3} & \pat{line4} & \pat{line5} & \pat{line6} & \phas{a} & \phas{b} & \psucc{a,b,next} \\\hline
     1 & 0 & 0 & 0 & 0 & 0 & 0 & 0 \\
     0 & 1 & 0 & 0 & 0 & 0 & 0 & 0 \\
     0 & 0 & 1 & 0 & 0 & 1 & 0 & 0 \\
     0 & 0 & 0 & 1 & 0 & 1 & 1 & 1 \\
     0 & 0 & 0 & 0 & 1 & 1 & 1 & 1 \\
     0 & 1 & 0 & 0 & 0 & 1 & 1 & 1 \\
     0 & 0 & 1 & 0 & 0 & 1 & 1 & 0 \\
     0 & 0 & 1 & 0 & 0 & 1 & 1 & 1
   \end{tabular}
   \caption{Predicate values for abstract thread objects}
   \label{tab:athreads}
\end{table*}

There are five thread locations, but eight possible distinct abstract thread objects,
as shown in Table~\ref{tab:athreads}. At the idle location, the fields \psvar{a} and
\psvar{b} are null, so the instrumentation predicates \phas{a}, \phas{b} and \psucc{a,b,next}
are false (0). They remain null/false at line 3, but at line 4, \psvar{a} is assigned a value
so \phas{a} is true (because \psvar{x} is also non-null). 
Then at line 5, \psvar{b} is assigned to \psvar{a}'s (non-null) successor, so \phas{b} and
\psucc{a,b,next} are true.
At line 6 after a successful CAS step, the fields are unchanged, so the instrumentation
predicates remain unchanged.
The fields are similarly unchanged after an unsuccessful CAS step, when the thread returns
to line 3; since the instrumentation predicates differ, threads at line 3 repeating the loop
are distinguished from those that have just started the operation.
Similarly at line 4, the instrumentation predicates distinguish initial from repeating
threads; in the latter case, the \psvar{a} field has been reassigned, so the \psucc{a,b,next}
instrumentation predicate may be false or may (by coincidence) still be true.

Knowing that there are eight distinct types of interleaving threads, we can predict
that there will be 255 abstract states --- the number of different possible combinations.
Indeed, that is the number reported by TVLA (see Table~\ref{tab:exresults}).

\subsection{Soft Invariants}
\label{sec:ex:si}

If we remove the location predicates and instrumentation predicates on thread fields
from the set of abstraction predicates, then all threads get mapped to a single abstract
thread object. The abstraction is too coarse though, and TVLA gives a spurious error for
our property --- the fact that \psvar{b} is non-null at line 5 is not preserved, so a
successful CAS step can result in \psvar{x} being set to null.

The properties we need to preserve are:
\begin{enumerate}
  \item At line 4, \phas{a} is true.
  \item At line 5, \phas{a}, \phas{b} and \psucc{a,b,next} are true.
\end{enumerate}
To do this, we define the following instrumentation predicates:
\[
  \pif{line4,\phas{a}}(v) \iff \pis{thread}(v) \land (\pat{line4}(v) \imp \phas{a}(v)) \\
  \pif{line5,\psucc{a,b,next}}(v) \iff \pis{thread}(v) \land (\pat{line5}(v) \imp \psucc{a,b,next}(v))
\]
We could also have defined similar instrumentation predicates \pif{line5,\phas{a}}
and \pif{line5,\phas{b}}, but \phas{a} and \phas{b} are both logical consequences of
\psucc{a,b,next}, so they are not necessary.

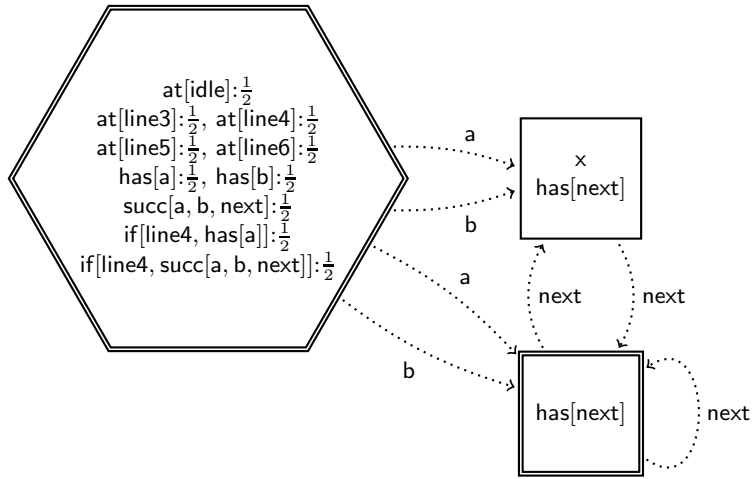
\begin{figure}
  \centering
  \begin{tikzpicture}
    \node[threads] (t) {\pat{idle}:\half\\
                        \pat{line3}:\half, 
                        \pat{line4}:\half\\
                        \pat{line5}:\half, 
                        \pat{line6}:\half\\
                        \phas{a}:\half, 
                        \phas{b}:\half\\
                        \psucc{a,b,next}:\half\\
                        \pif{line4,\phas{a}}:\half\\
                        \pif{line4,\psucc{a,b,next}}:\half};
    \node[nodeCA,right=of t] (nx) {\px\\\phas{next}};
    \node[nodesCA,below=of nx] (ns) {\phas{next}};

    \path (t) edge [maybe,bend left=10] node {\pa} (nx)
          (t) edge [maybe,bend right=10] node [swap] {\pb} (nx)
          (t) edge [maybe,bend left=10] node {\pa} (ns)
          (t) edge [maybe,bend right=10] node [swap] {\pb} (ns)
          (nx) edge [maybe,bend left] node {\pnext} (ns)
          (ns) edge [maybe,loop right] node {\pnext} (ns)
          (ns) edge [maybe,bend left] node [swap] {\pnext} (nx);
  \end{tikzpicture}
  \caption{Single abstract state}
  \label{fig:one}
\end{figure}

With these two soft invariant instrumentation predicates,
TVLA constructs an abstract system that verifies our property, and has just
one abstract state --- shown in Figure~\ref{fig:one} ---
two orders of magnitude less than without soft invariants.
Table~\ref{tab:exresults} shows the results reported by TVLA with and without
soft invariants, for unbounded ($\infty$) threads.

\begin{table}
  \centering
  \begin{tabular}{ccr|rrrrr}
       &            &  Heap &       &  Ave  &   Max &        \\
       &            & Limit &  Time &  RAM  &   RAM & CA & Stored \\
   SIs & Th.        &  (MB) &   (s) &  (MB) &  (MB) & States & States \\\hline
     N & \(\infty\) &   800 &   6.1 &   131 &   279 & 255 &  4,335  \\
     Y & \(\infty\) &   800 &   0.3 &     1 &     1 &   1 &      7  \\
  \end{tabular}
  \caption{Example verification results}
  \label{tab:exresults}
\end{table}

\subsection{Safety}
\label{sec:ex:safe}

In this example, we chose two soft invariants, which were both
necessary and sufficient to verify the property of interest. We could 
have defined further soft invariant predicates though, without affecting the
safety of the analysis.
There are three possible cases: properties that:
\begin{enumerate}
\item are invariant and needed, but are logical consequences of others specified;
\item are invariant but not needed; or
\item are not invariant.
\end{enumerate}

For the first case, we previously mentioned the possibility of
defining the predicates \pif{line5,\phas{a}} and \pif{line5,\phas{b}}.
For the second case, we can look at Table~\ref{tab:athreads}
and see the possibility of defining, e.g.
\[
  \pif{idle,\phas{a}}(v) \iff \pis{thread}(v) \land (\pat{idle}(v) \imp \lnot\phas{a}(v)) \\
  \pif{line6,\psucc{a,b,next}}(v) \iff \pis{thread}(v) \land (\pat{line6}(v) \imp \psucc{a,b,next}(v))
\]
However, these properties, whilst invariant, do not impact on the
verification of the property, so can be left out to avoid the overhead
they would incur.
For the third case, we could for example ``mistakenly'' define
\[
  \pif{line3,\phas{a}}(v) \iff \pis{thread}(v) \land (\pat{line3}(v) \imp \lnot\phas{a}(v))
\]
This property is not invariant because of the loop, but defining it does
not assert the property in the system. When a thread performs an unsuccessful CAS
step at line 5, the resulting thread object would have
\pif{line3,\phas{a}} false. Thus the single abstract thread object would
have this soft invariant unknown.
There is some overhead incurred maintaining the non-invariant predicate,
but the analysis is otherwise unaffected.

See Section~\ref{sec:choice} for more discussion on this topic.

\section{Application}
\label{sec:results}

In previous work \cite{frig:shap14} we have used canonical abstraction
to (attempt to) verify linearizability \cite{herl:line90} for three
nonblocking concurrent data structure algorithms. We had mixed success,
due to the size of the abstract statespaces exceeding memory resources.
Here we demonstrate greatly improved results using soft invariants.

\subsection{Stack}
\label{sec:results:stack}

The first algorithm is a stack, introduced by Treiber \cite{trei:syst86}.
Versions of the algorithm have been formally verified by 
many authors \cite{colv:veri05,berd:thre08,frig:shap14};
our analysis adds to this and demonstrates the applicability of canonical abstraction.

\subsubsection{Base Model}

The model is described in detail elsewhere \cite{frig:mode13,frig:shap14}
but contains 30 core predicates, including 16 location predicates,
and 19 instrumentation predicates.
Because TVLA does not implement partial order reduction, we constructed two
models --- one with manually specified restrictions on thread interleaving at
certain locations, and one with full interleaving.
We were able to verify linearizability for the restricted interleaving model
with unbounded threads, list nodes and data values. For the model with full
interleaving, we were only able to verify linearizability with three threads ---
with four or more threads the statespace was too large.

\subsubsection{Soft Invariants}

To identify the necessary and sufficient soft invariants, we used the same
approach we took to identify instrumentation predicates in the initial model --- a
combination of inference and trial-and-error. We constructed tables equivalent
to Table~\ref{tab:athreads} for the example, and for the restricted model we were
able to check this against the abstract statespace reported by TVLA.

We defined 12 soft invariants for the restricted interleaving model and 22
for the full interleaving model --- the lists and full discussion are
omitted for space reasons, but are available elsewhere \cite[Chapter 8]{frig:mode13}.

There are two notable points to be made, relating to more complex properties.
First, there were cases where the value of a predicate at a location depended
on the value of another predicate, so we defined conditional soft invariants
with structures like:
\[
  \pif{X,Q}(v) \iff \pis{thread}(v) \land (\pat{X}(v) \imp (\pp{P}(v) \imp \pp{Q}(v))) \\
  \pif{X,R}(v) \iff \pis{thread}(v) \land (\pat{X}(v) \imp (\lnot\pp{P}(v) \imp \pp{R}(v)))
\]
Second, there was a more complicated property that was most easily addressed
by defining a new instrumentation predicate, and then defining a soft invariant
for the value of that new predicate.
Adding another instrumentation predicate to the model was extra work, but was
an extension of the work involved in identifying the instrumentation predicates
in the inital model.

\subsubsection{Results}

We analysed the models using TVLA 3.0\(\alpha\) on a machine with an Intel 
Core~2 3.0\,GHz processor and 4\,GB of RAM, running Java 1.6.0 on
a 32-bit GNU/Linux operating system.
Tables \ref{tab:stack:res} and \ref{tab:stack:full} show the results of
verifying linearizability for both models, with and without soft invariants.
The first three columns note whether soft invariants were used, the bound
on the number of threads and how much RAM was
allocated to TVLA; the remainder are numbers reported by TVLA's output.

Collapsing threads using soft invariants is very effective, with a 99.2\% reduction
in the abstract statespace, and a
99.4\% reduction in
states stored, for the restricted interleaving model.

\begin{table}
  \centering
  \begin{tabular}{ccr|rrrrr}
       &            &  Heap &       & Ave &   Max &    \\
       &            & Limit &  Time & RAM &   RAM &   CA   & Stored   \\
   SIs & Th.        &  (MB) &   (s) & (MB)&  (MB) & States & States \\\hline
     N &    3       &   800 &   102 & 173 &   336 &    440 &  6,493 \\
     N & \(\infty\) & 2,048 & 1,934 & 849 & 1,603 &  1,910 & 74,056 \\
     Y & \(\infty\) &   800 &    25 & 115 &   280 &     16 &    447 \\
  \end{tabular}
  \caption{Stack verification results with restricted interleaving}
  \label{tab:stack:res}
\end{table}

The collapsed threads approach is even more effective for the full interleaving model, 
with the collapsed unbounded statespace
99.8\% reduced compared to the non-collapsed 3-thread statespace. This indicates that the
reduction compared to the non-collapsed unbounded statespace should be far greater.

Compared to the results in Table~\ref{tab:stack:res}, the collapsed unbounded model stores
more than twice the number of states, a great improvement over the non-collapsed 3-thread model, which
stores nearly 23 times as many states.
The amount of memory used by TVLA only increases slightly, but the analysis takes
nearly six times as long.
However, the analysis is still 13 times faster than the model \emph{with}
restricted interleaving and \emph{without} collapsed threads.

\begin{table}
  \centering
  \begin{tabular}{ccr|rrrrr}
       &            &  Heap &       &  Ave  &   Max &    \\
       &            & Limit &  Time &  RAM  &   RAM & CA & Stored   \\
   SIs & Th.        &  (MB) &   (s) &  (MB) &  (MB) & States & States \\\hline
     N &   3        & 2,048 & 2,931 & 1,089 & 1,946 & 21,107 & 148,191 \\
     Y & \(\infty\) &   800 &   141 &   149 &   295 &     32 &     932 \\
  \end{tabular}
  \caption{Stack verification results with full interleaving}
  \label{tab:stack:full}
\end{table}

\subsection{Queues}
\label{sec:results:queues}

The remaining algorithms are two nonblocking concurrent 
queue algorithms, the original introduced by Michael and Scott \cite{mich:nonb98} 
(``MS queue''), and a modified version by Doherty et~al.~\cite{dohe:form04} (``DGLM queue'');
the latter authors provided the first formal verification of linearizability.

\subsubsection{Base Models}

These algorithms have a similar singly-linked-list structure,
so construction of the canonical abstraction models was relatively straightforward ---
most of the core and instrumentation predicates were able to be copied from the stack
model. 
The models were larger, with 44 core predicates, including 26 thread locations,
and 22 instrumentation predicates.

As with the stack, we constructed two models for each queue --- one with manually
restricted interleaving and one with full interleaving.

Previously \cite{frig:shap14}, we were able to verify linearizability for the restricted
interleaving models with two threads, but not three.
For the models with full interleaving, even when bounded to two threads the statespace was
too large for the memory available.

\subsubsection{Soft Invariants}

We followed a similar process as for the stack, and defined 25--57 soft invariants.
The lists and full discussion are omitted for space reasons, but are
available elsewhere \cite[Chapter 8]{frig:mode13}.
Table~\ref{tab:queue:locsi} shows the number of distinct interleaving thread configurations
and the number of defined soft invariants for each component of the two algorithms
 --- idle state (in between operations), enqueue operation (shared by both), the MS and DGLM dequeue
operations, and then the two algorithms in total.

\begin{table}
  \centering
  \begin{tabular}{c|rr|rr}
    & \multicolumn{2}{c}{Restricted} & \multicolumn{2}{c}{Full} \\
    & locs & SIs & locs & SIs \\ \hline
    idle & 1 & 1 & 1 & 1 \\
    enqueue & 5 & 11 & 10 & 18 \\
    dequeue$_M$ & 5 & 23 & 12 & 38 \\
    dequeue$_D$ & 5 & 13 & 12 & 24 \\ \hline
    MS & 11 & 35 & 23 & 57 \\
    DGLM & 11 & 25 & 23 & 43
  \end{tabular}
  \caption{Interleaving location and soft invariant counts for queue models}
  \label{tab:queue:locsi}
\end{table}

As with the stack, we defined conditional soft invariants.
We also had to define a new instrumentation predicate for the MS queue, but for
a different property than the stack.

\subsubsection{Results}

For both the restricted and full interleaving models, we were able to
verify linearizability for unbounded threads, list nodes and data values,
with relatively small resources --- a considerable improvement over the
models without soft invariants.
Tables \ref{tab:queue:res} and \ref{tab:queue:full} show the results.

\begin{table*}
  \centering
  \begin{tabular}{cccr|rrrrr}
       &     &            &  Heap &      &  Ave  &  Max &    \\
       &     &            & Limit & Time &  RAM  &  RAM & CA & Stored \\
   Deq & SIs &    Th.     &  (MB) &  (s) &  (MB) & (MB) & States & States \\\hline
   MS  &  N  &     2      &   800 &  393 &   260 &  476 & 3,770 & 24,271 \\
   MS  &  Y  & \(\infty\) &   800 &   81 &   151 &  309 &    12 &    523 \\
  DGLM &  N  &     2      &   800 &   83 &   189 &  354 & 1,506 & 10,746 \\
  DGLM &  Y  & \(\infty\) &   800 &  216 &   181 &  310 &    14 &    591  
  \end{tabular}
  \caption{Queue verification results with restricted interleaving}
  \label{tab:queue:res}
\end{table*}

\begin{table*}
  \centering
  \begin{tabular}{cccr|rrrrr}
       &     &            &  Heap &       &  Ave  &   Max &    \\
       &     &            & Limit &  Time &  RAM  &   RAM & CA & Stored   \\
   Deq & SIs &    Th.     &  (MB) &   (s) &  (MB) &  (MB) & States & States \\\hline
   MS  &  Y  & \(\infty\) &   800 & 3,109 &   304 &   727 &    96  &  4,148 \\
  DGLM &  Y  & \(\infty\) &   800 & 2,887 &   290 &   731 &   112  &  4,604 \\
  \end{tabular}
  \caption{Queue verification results with full interleaving}
  \label{tab:queue:full}
\end{table*}

An interesting observation is that for the statespaces of the non-collapsed models,
the MS queue is much larger --- 2.5 times larger with a bound of two threads.
For the statespaces of the collapsed models though, the DGLM queue is larger,
by 17\%.
The DGLM queue allows the Head and Tail pointers to
``cross'' \cite{dohe:form04}, unlike the MS queue, which results in more canonically abstract list
configurations, and accounts for the collapsed model differences. In the non-collapsed
models, the MS queue has more canonical thread objects --- which is related to the larger number of
soft invariants needed --- and this becomes a larger factor than the differing number
of list structures.

\section{Choosing Soft Invariants}
\label{sec:choice}

Collapsing thread objects with soft invariants provides an exponential reduction in the size of the
statespace, as evidenced by the results presented so far.
Extending models with soft invariants requires extra manual effort though, so some discussion
on the subject is warranted.

Ultimately, soft invariants do not add anything technically new to canonical abstraction ---
they are ``just'' instrumentation predicates --- so their discovery and definition
is an extension of the effort necessary for these anyway.
In the basic case, a spurious error is reported, which indicates that the abstraction is too
coarse. Examining the output, the user can infer what property is needed to be preserved,
and then define an instrumentation predicate to refine the abstraction.

For the models we have looked at, defining soft invariants was relatively little extra effort,
as they mostly built on the effort of determining the thread instrumentation predicates.
Consider the example system from Section~\ref{sec:ex} --- the two soft invariants require
an understanding of the values and mutual relationship of the \psvar{a} and \psvar{b} fields
at lines 4 and 5. But those facts had already been determined in order to define the
instrumentation predicates \phas{a}, \phas{b} and \psucc{a,b,next}.
Constructing a model with collapsed threads initially --- as opposed to
extending an existing model --- may make determining soft invariants less straightforward.
In these cases we expect that over defining the soft invariants may be effective.

\subsection{Over Defining Soft Invariants}
\label{sec:sec:choice:over}

The amount of effort needed to identify soft invariants could be mitigated, especially
for unfamiliar systems, by exploiting the ``soft'' nature of soft invariants and
over defining them. As mentioned in Section~\ref{sec:ex:safe}, if a property is not
actually invariant, there is no consequence for the abstract statespace, it just adds
some overhead to the analysis.
Thus we can define a range of soft invariants, many of which may not be necessary
or even correct.

To confirm and quantify this idea, we modified the stack model with full interleaving
from Section~\ref{sec:results:stack} with soft invariants for every thread predicate.
For every thread predicate \ppp, and every location \emph{loc}, we defined two
soft invariants:
\[
  \pif{loc,\ppp}(v) \iff \pis{thread}(v) \land (\pat{loc}(v) \imp \ppp(v)) \\
  \pnif{loc,\ppp}(v) \iff \pis{thread}(v) \land (\pat{loc}(v) \imp \lnot\ppp(v)) \\
\]
We kept the six conditional soft invariants, distinguishing them with a `c' prefix.

Table~\ref{tab:stack:si} shows the results of the analysis, with the first line
repeated from Table~\ref{tab:stack:full}.
The canonically abstract statespace is unchanged with 32 states, though fewer
intermediate states are stored.
Memory use increases and the time roughly doubles.

\begin{table}
  \centering
  \begin{tabular}{cr|rrrrr}
       &  Heap &       &  Ave  &   Max &        \\
   Over& Limit &  Time &  RAM  &   RAM &   CA   & Stored \\
   def.&  (MB) &   (s) &  (MB) &  (MB) & States & States \\\hline
     N &   800 &   141 &   149 &   295 &     32 &   932  \\
     Y &   800 &   278 &   202 &   338 &     32 &   726  \\
  \end{tabular}
  \caption{Stack verification results with extra soft invariants}
  \label{tab:stack:si}
\end{table}

These results show that the tradeoff between analysis efficiency and
determining a minimal set of soft invariants is likely to be a practical one.
Here, at least half of the properties defined are not invariant, with some being
false in the initial state, and the increases in time and memory are acceptable.

\section{Related Work}
\label{sec:relwork}

The closest work to ours is by
Manevich et~al.~\cite{mane:heap08} and
Berdine et~al.~\cite{berd:thre08},
who also attempt to address the state explosion problem caused by
canonical thread objects.
Their approaches are orthogonal to ours, as they attempt to improve the
storage of states by decomposing each state in to thread components and
storing each unique component only once. They verify the same algorithms
that we do, but it is difficult to directly compare results, as they extend
previous work \cite{amit:comp07} that uses an additional technique (delta
heap abstraction) to abstract lists for verifying linearizability, and which
does not abstract threads; we extend previous work \cite{frig:shap14} that
verifies linearizability for unbounded threads using canonical abstraction alone.
Nonetheless, their approach reduces, but does not remove, the effect of
canonical thread permutations, so our approach may scale better for larger
systems.


\section{Conclusions}
\label{sec:conc}

We have presented a novel approach to canonical abstraction, which exponentially
reduces the statespace size by collapsing all thread objects in a state into a single summary
object. This has allowed us to verify linearizability for three concurrent data structure
algorithms with unbounded threads, list lengths and data values, 
where previously the statespaces were impractically large.

The coarse abstraction on threads is refined by defining ``soft invariant'' instrumentation
predicates, which preserve properties of threads at specific locations. The identification
of sufficient soft invariants is currently a manual process. However, the soft invariants are
safe to define --- correctness is not affected if the properties are not actually invariant ---
so they can be over-defined to reduce the manual effort. Identifying the minimal set, or even
a likely superset, may well be a problem amenable to automation.

This approach should be applicable to any multi-process canonical abstraction model. In
fact, there is nothing inherently process/thread-specific about these instrumentation
predicates, so the same idea could be applied to other types of models to reduce the
abstract statespace.

\bibliographystyle{splncs03}
\bibliography{fg14collapse}

\end{document}